\documentclass[letterpaper,english,aps,prl,twocolumn,floatfix,showpacs,amsfonts,amssymb,superscriptaddress]{revtex4}

\usepackage[T1]{fontenc}
\usepackage[latin1]{inputenc}
\usepackage{graphics}
\usepackage{amssymb}
\usepackage{amsmath}
\usepackage{epsfig}

\newcommand{\beq}{\begin{equation}}
\newcommand{\eeq}{\end{equation}}
\newcommand{\bea}{\begin{eqnarray}}
\newcommand{\eea}{\end{eqnarray}}

\newcommand{\rmd}{{\rm d}}

\newcommand{\cuoo}{CuO$_{2}$}

\newcommand{\lasco}{La$_{2-x}$Sr$_{x}$CuO$_{4}$}

\begin{document}
\def\tende#1{\,\vtop{\ialign{##\crcr\rightarrowfill\crcr
\noalign{\kern-1pt\nointerlineskip} \hskip3.pt${\scriptstyle
#1}$\hskip3.pt\crcr}}\,}

\title{Lightly doped {\lasco} as a Lifshitz helimagnet}

\author{V.\ Juricic}

\affiliation{Institute for Theoretical Physics, University of
Utrecht, Leuvenlaan 4, 3584 CE Utrecht, The Netherlands.}

\author{M.\ B.\ Silva Neto}

\affiliation{Institute for Theoretical Physics, University of
Utrecht, Leuvenlaan 4, 3584 CE Utrecht, The Netherlands.}

\author{C.\ Morais Smith}

\affiliation{Institute for Theoretical Physics, University of
Utrecht, Leuvenlaan 4, 3584 CE Utrecht, The Netherlands.}

\begin{abstract}
We study the static magnetic correlations in lightly doped {\lasco}
within the framework of a dipolar frustration model for a canted
antiferromagnet. We show that the stability of the canted N\' eel
state for $x<2\%$ is due to the Dzyaloshinskii-Moriya and XY
anisotropies. For higher doping the ground state is unstable towards
a {\it helicoidal} magnetic phase where the transverse components of
the staggered magnetization rotate in a plane perpendicular to the
orthorhombic $b$-axis. Our theory reconciles, for the first time, the
incommensurate peaks observed in elastic neutron scattering with Raman
and magnetic susceptibility experiments in {\lasco}.
\end{abstract}

\maketitle

{\it Introduction}.$-$
The magnetic and transport properties of {\lasco} (LSCO) evolve
tremendously with doping \cite{Kastner}. For $0<x<x_{AF}\simeq 0.02$,
LSCO is a Mott-Hubbard insulator that exhibits 3D antiferromagnetic (AF)
order with a N\' eel temperature $T_N\simeq 325$K at $x=0$. For
$x_{AF}<x<x_{SC}\simeq 0.055$, inside the spin-glass (SG) phase,
static incommensurate (IC) magnetic correlations are observed in
neutron scattering, demonstrating the importance and persistence
of the strong AF correlations well after the N\'eel long range
order has been destroyed. Further doping, $x>x_{SC}$, eventually
drives the system into a superconducting (SC) state, where dynamic
magnetic correlations are also known to play an important role
\cite{Kastner}.

Inelastic neutron scattering (INS) experiments within the SC phase
have revealed that dynamical IC spin correlations coexist with
superconductivity \cite{yamada}. Moreover, the experimentally
observed four peaks associated with the IC magnetic order are
accompanied by charge peaks, with periodicity twice the magnetic
one, provided a low-temperature tetragonal phase is stabilized (as,
for example, by Nd doping) \cite{tranq}. These features have been
consistently interpreted within a model of stripes acting as antiphase
domain walls \cite{zaanen}. More recently, static IC magnetic
correlations have also been observed within the low-temperature
orthorhombic (LTO) SG phase of LSCO \cite{NSSG}. In this case,
only two IC peaks, rotated by 45$^\circ$ with respect to the ones
in the SC phase, have been observed. Because such
incommensurability follows the same linear dependence upon doping
as the other four peaks \cite{matsuda}, the above results, as well as
recent Raman measurements \cite{Tassini}, have been interpreted as a
signature of diagonal stripe order. However, associated charge peaks
have never been observed in the LTO phase of LSCO, what raises the
question of the validity of the diagonal stripe picture
within the insulating SG phase.

Recently, a model based on the seminal ideas of Shraiman and
Siggia \cite{SS1} has been proposed. It accounts for the observed
magnetic IC diagonal peaks without having to assume a charge order
\cite{nils}. Within this model, the two IC magnetic peaks would
originate from a spiral state, with the staggered magnetization
rotating within the {\cuoo} layers ($ab$ plane), in such a way
that the measured magnetic incommensurability would correspond to
the inverse spiral pitch. More recently, Sushkov and Kotov have
considered a similar spiral state in the
$t-t^\prime-t^{\prime\prime}-J$ model \cite{SK}. It was shown that
while in the insulating region, $x<x_{SC}$, the diagonal $(1,1)$
and $(1,-1)$ spirals have lower energy than the horizontal $(1,0)$
and vertical $(0,1)$ ones due to the Coulomb trapping of the doped
hole near the Sr ion, in the metallic state, $x>x_{SC}$, the Fermi
motion energy favors the $(1,0)$ and $(0,1)$ spiral states,
leading to a jump of 45$^\circ$ in the direction of the spiral
pitch across the metal-insulator transition.

Despite being able to qualitatively explain neutron scattering
data in the SG phase, the above spiral pictures have two major
problems: i) the collinear N\'eel state is unstable towards the
spiral state already at infinitesimal doping, $x\neq 0$; ii) the
spiraling of the staggered magnetization in the {\cuoo} layers is
not consistent with magnetic susceptibility experiments in LSCO,
which indicate that the Cu$^{++}$ spins remain confined to the
$bc$ plane
throughout the SG phase ($\chi_a$ remains featureless)
\cite{Lavrov}. In this Letter we demonstrate that only by
considering properly both the Dzyaloshinskii-Moriya (DM) and XY
anisotropies the above two problems can be solved, and we provide
an elegant description of the unified physics of IC correlations
and anisotropic magnetic response in the SG phase of LSCO. We show
that the anisotropies give {\it robustness} to the {\it canted}
N\'eel state for $x<x_{AF}$ and that the same anisotropies lead to
the formation of a {\it Lifshitz helimagnet} above $x_{AF}$, with
the small transverse components of the staggered magnetization
rotating in the $ac$ plane. Such helix rotation of the
staggered magnetization is accompanied by a small precession of
the {\it local} magnetization (weak-ferromagnetic (WF) moment) around
the out-of-plane $c$-axis. The {\it space integrated} sublattice
magnetization and WF moment, on the other hand,
are oriented along the $b$- and $c$-axis, respectively, in such a
way that, from the point of view of the magnetic susceptibility,
the total spin is confined to the $bc$ plane, in agreement with
experiments \cite{Lavrov}.

{\it The model}.$-$
The presence of static magnetic correlations within the SG phase of
LSCO allows us to use a Hamiltonian description of the long-wavelength
fluctuations of the {\it staggered} order parameter, ${\bf n}$ (see
\cite{Papanicolaou,marcello} for a detailed derivation of the
model). For $x=0$ we can write (we use a soft version of the
${\bf n}^2=1$ constraint)
\begin{equation}\label{LS}
{\cal H}_M=\frac{1}{2t}\int d^2{\bf x}\left \{(\nabla{\bf n})^2+
\frac{1}{2}\left(\frac{m_\alpha}{Ja}\right)^2 n_\alpha^2
+\frac{u_0}{2}({\bf n}^2-1)^2\right\},
\end{equation}
where ${\bf n}$ is the local staggered magnetization,
$t^{-1}\equiv \rho_s/T$ is the renormalized classical spin
stiffness, with $\rho_s=JS^2$, $J$ is the AF superexchange, $a$ is
the lattice constant, $T$ is the temperature, and $S=1/2$ is the
spin. The effect of both the DM and XY anisotropies is to generate
gaped transverse excitations. We define $m_\alpha$, ($\alpha$
$\equiv a,b,c$), as the bare masses of the magnetic excitations
along the three directions of the LTO phase. The bare masses $m_a$
and $m_c$ are related to the DM and XY anisotropy parameters,
respectively, whereas the mass of the longitudinal mode, $m_b$
(the heaviest of all three), is proportional to the coupling
constant $u_0$. Furthermore, the {\it uniform} part of the
Cu$^{++}$ spins, ${\bf L}$, is related to the staggered order
parameter, ${\bf n}$, through $\langle{\bf
L}\rangle=(1/2)(\langle{\bf n} \rangle\times{\bf D}_+)$, and gives
rise to a WF moment perpendicular to the {\cuoo}
layer \cite{Papanicolaou,marcello}. Here ${\bf D}_+=(D_+,0,0)$ is
a thermodynamic DM vector oriented along the $a$-axis, related to
the tilting angle of the oxygen octahedra $D_+\sim\delta$
\cite{marcello}, and we use units such that $J=a=1$.

The effect of a small number of holes ($x\neq 0$) on an
antiferromagnet can be characterized, in a coarse-grained description,
in terms of a dipolar frustration of the AF background \cite{SS1,nils}.
We introduce a dipolar field, ${\bf P}_{\mu}$, that couples to the AF
magnetization current as
\begin{equation}
\label{inter}
{\cal H}_{int}=-2\lambda\int  d^2{\bf x}\,{\bf P}_\mu\cdot({\bf
n}\times\partial_\mu{\bf n}),
\end{equation}
where $\lambda\equiv{\tilde \lambda}/T$ is the dipolar coupling
constant, ${\tilde\lambda}\sim 1$ \cite{SS}, and the field
$P^\alpha_\mu$ is a vector in both lattice and spin spaces. The above
mathematical form belongs to a class known as {\it Lifshitz invariants},
which play an important role in stabilizing long-period
spatially modulated structures with fixed sense of rotation of the
vectors ${\bf n}$ in systems like Ba$_2$CuGe$_2$O$_7$ and
K$_2$V$_3$O$_8$ \cite{bogdanov}.


In the case of a nonuniform configuration for the dipolar field,
${\bf P}_{\mu}$, we can write
\begin{equation}\label{dipoles}
{\cal H}_D=\frac{1}{2\kappa}\int d^2{\bf x}\left\{ (\nabla{\bf
P}_\mu)^2+\mu_{\alpha\mu}^2(P^\alpha_\mu)^2\right\},
\end{equation}
where $\kappa={\tilde \kappa}T$ with ${\tilde\kappa}$ denoting the
dipole stiffness, and $\mu_{\alpha\mu}$ is the bare mass of the dipole
field, $P_\mu^\alpha$, which, in microscopic terms, is related to the
energy cost of populating the $\mu$th valley of the vacancy Fermi
surface with the spin polarization $\alpha$ \cite{SS1,SS}.

The appearance of elastic IC peaks in neutron scattering can
occur, in principle, whenever $\langle{\bf P}_\mu\rangle\neq 0$.
We thus calculate the effect of the magnons on the self-energy of
the dipolar fields, arising from the dipole-magnon interaction in
Eq. (\ref{inter}). Straightforward calculations yield
\begin{eqnarray}
\label{massredd}
{\cal M}^2_{\alpha\mu}&=& \mu^2_{\alpha\mu}- 8\kappa(t\lambda)^2
\epsilon_{\alpha\beta\gamma}\epsilon_{\alpha\beta\gamma}\nonumber\\
&\times& \int\frac{d^2{\bf k }}{(2\pi)^2}\frac{k_\mu
k_\mu}{(k^2+m_\beta^2/2)(k^2+m_\gamma^2/2)}.
\end{eqnarray}
The Fermi wavevector $k_F=\sqrt{\pi x}$ provides the cutoff in the
momentum integrals, because the coarse-grained description employed
here is valid on length scales much larger than the distance
between the holes, $l\gg k_F^{-1}\sim x^{-1/2}$ \cite{SS1,SS}.
Inspection of Eq.\ (\ref{massredd}) reveals that, even when the bare
dipole masses are isotropic in the spin space, $\mu_{\alpha\mu}
\equiv \mu_{\mu}$, $\forall \alpha$, the renormalized ones may be
anisotropic due to the dipolar interaction, Eq.\ (\ref{inter}),
which involves different magnon modes. Moreover, an instability to a
phase with $\langle P_\mu^\alpha\rangle\neq 0$ may indeed occur.
In particular, when $m_a=m_c=0$, such instability appears already
at infinitesimal doping, because the momentum integral in
Eq.\ (\ref{massredd}) diverges logarithmically. Finally, due to the
hierarchy between the masses of the magnon modes, $m_a<m_c\ll m_b$
(actually $m_b\rightarrow\infty$ in the nonlinear ${\bf n}^2=1$
description), we find that ${\cal M}^2_{\alpha\mu}$ will become
negative first for the spin component of the dipolar field along
the orthorhombic $b$-axis.

{\it Canted N\'eel state}.$-$
Let us consider first the low doping regime where $\langle
P_\mu^\alpha\rangle=0$. In this case, the magnetism is commensurate
and the fluctuating dipoles give rise to corrections to the magnon
self-energy in one-loop approximation. The renormalized propagator
for the $n_\alpha$ field reads $G_{\alpha\beta}({\bf q})=
t\delta_{\alpha\beta}(q^2+m_\alpha^2/2-\Sigma_\alpha({\bf q}))^{-1}$,
where ($\alpha\neq\beta\neq\gamma$)
\begin{eqnarray}\label{magnonselfen}
\Sigma_\alpha({\bf
q})&=&\kappa(t\lambda)^2\epsilon_{\alpha\beta\gamma}
\epsilon_{\alpha\beta\gamma}\nonumber\\
&\times&\int \frac{d^2{\bf k}}{(2\pi)^2}
\frac{(k+q)_\mu(k+q)_\mu}{(k^2+m_\beta^2/2) [({\bf k}-{\bf
q})^2+\mu_{\gamma\mu}^2]}.
\nonumber
\end{eqnarray}
The above self-energy is an analytical function of the momentum
and thus we write $ \Sigma_\alpha ({\bf q})=\Sigma_\alpha(0)+(1/2)
q_\mu q_\nu\partial^2\Sigma_\alpha({\bf q})/\partial q_\mu\partial
q_\nu\vert_{{\bf q}=0}$. The first term in the self-energy,
$\Sigma_\alpha(0)$, gives rise to corrections to the magnon mass
\begin{eqnarray}\label{massred}
M^2_\alpha &=& m^2_\alpha-2\kappa(t\lambda)^2
\epsilon_{\alpha\beta\gamma}\epsilon_{\alpha\beta\gamma}\nonumber\\
&\times& \int\frac{d^2{\bf k }}{(2\pi)^2}\frac{k_\mu
k_\mu}{(k^2+m_\beta^2/2)(k^2+\mu_{\gamma\mu}^2)},
\end{eqnarray}
whereas the second term leads to a reduction of the spin
stiffness,
\begin{eqnarray}\label{stiffness}
{\tilde\rho}_{s\mu\nu}&=&\rho_s\delta_{\mu\nu}-
\delta_{\mu\nu}\kappa(t\lambda)^2\epsilon_{\alpha\beta\gamma}
\epsilon_{\alpha\beta\gamma}\nonumber\\
&\times&\int\frac{d^2{\bf k}}{(2\pi)^2}
\frac{(k^2)^2+\mu_{\gamma\nu}^4}
{(k^2+m_\beta^2/2)(k^2+\mu_{\gamma\nu}^2)^3}.
\end{eqnarray}

\begin{figure}[htb]
\begin{center}
\includegraphics[width=6cm,angle=-90]{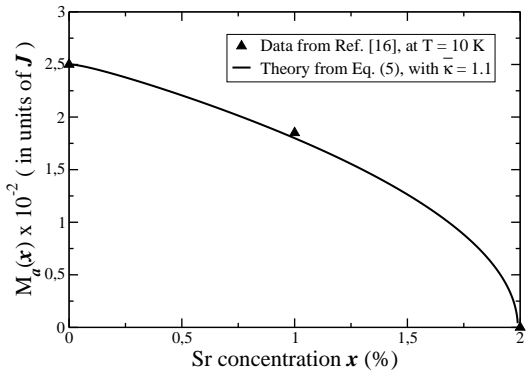}
\caption{Doping dependence of the DM gap as given by Eq.\
(\ref{massred}) using the bare parameters of the model.
Experimental data from Ref. \cite{gonzar} taken at $T=10$ K.}
\label{Fig-DM-Gap}
\end{center}
\end{figure}

It is clear now from Eqs. (\ref{massredd}) and (\ref{massred})
that it is exactly the DM and XY anisotropies that give {\it
robustness} to the canted N\'eel state at low doping, because they
lead to nonvanishing magnon gaps, $m_a\neq 0$ and $m_c\neq 0$.
Most remarkably, we find that the reduction of the DM gap, $M_a$,
is consistent with recent Raman scattering experiments by Gozar
{\it et al.} \cite{gonzar}, which show that the DM gap decreases
with doping and vanishes at $x=x_{AF}\simeq 2\%$, when the SG
phase sets in. Experimentally, at $x=1\%$ the decrease of the DM
gap is about 26$\%$ \cite{gonzar}. Such huge renormalization,
however, {\it cannot} be explained simply from the decrease of the
tilting angle of the oxygen octahedra in the LTO phase of LSCO,
which is known to happen with Sr doping. Our model yields the
decrease of the DM gap with doping due to the coupling of the
background magnetization current to the dipoles, Eq.
(\ref{inter}). Using Eq.\ (\ref{massred}) with the bare values of
parameters, ${\tilde \kappa}\sim1.1$, ${\tilde\lambda}\sim 1$, $t\sim
1/S^2$, $m_a\sim 2.5\cdot 10^{-2}$, $m_c\sim 5\cdot 10^{-2}$, and
$\mu\sim 1$ \cite{SS}, we determine the doping dependence of the
DM gap, at $T=10$ K (see Fig. \ref{Fig-DM-Gap}). The reduction of
the DM gap at $x=1\%$ is about 27$\%$, in remarkable agreement with
the experiments. Moreover, from the vanishing of the DM gap we
obtained the critical concentration $x^{th}_{AF}=0.0198\simeq 2\%$, as
experimentally observed \cite{gonzar}, signaling the instability of
the N\' eel phase.

%

Our results are also consistent with recent magnetic susceptibility
measurements \cite{Lavrov}. Within our model, the only effect of
doping is the reduction of the magnon masses and of the spin stiffness
(see Eqs. (\ref{massred}) and (\ref{stiffness})). Therefore, we expect
the qualitative features of the susceptibility in the AF region of LSCO
to remain the same as in the undoped compound \cite{marcello}. The peaks
in both $\chi_b$ and $\chi_c$ susceptibilities will be shifted towards
lower temperatures, because both the reduction of the magnon gaps and
of the stiffness with doping lead to a decrease of the N\' eel temperature,
$T_N$. This is indeed observed experimentally \cite{Lavrov}. Furthermore,
we find that the unusual $T=0$ hierarchy of the susceptibilities,
$\chi_a<\chi_c<\chi_b$, is preserved \cite{marcello} (once the
anisotropic van Vleck contribution to $\chi_c$ is subtracted
\cite{Lavrov}) with $\chi_a\approx \sigma_0^2/t$,
$\chi_b=(1/t)(D_+^2/M_{c}^2(x))$, and
$\chi_c\approx\chi_a+(1/t)(D_+^2/M_{b}^2(x))$,
where $\sigma_0$ measures the Cu$^{++}$ effective moment \cite{marcello}.
%
%

\begin{figure}[htb]
\begin{center}
\includegraphics[width=6cm]{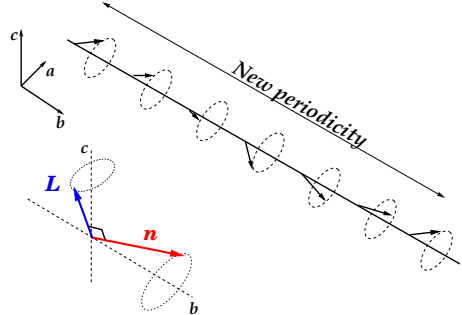}
\caption{(Color online) Helix rotation of the staggered magnetization for
  $x>x_{AF}$. Notice that the longitudinal component is the largest,
  and the small transverse ones describe an ellipse in the $ac$ plane.}
\label{Fig-helicoid}
\end{center}
\end{figure}

{\it Helicoidal state}.$-$
Let us consider now the case of higher doping where an instability
towards a phase with $\langle P_\mu^\alpha\rangle\neq 0$ may occur.
The easy axis for the spin part of the dipolar field is the orthorhombic
$b$-axis, which is a consequence of the hierarchy $m_a<m_c\ll m_b$.
As the lattice part is concerned, our model is isotropic, and, as
such, cannot determine the lattice direction of the dipolar field.
Motivated by neutron scattering experiments, we assume that the
dipolar field acquires the expectation value along the $b$-axis in
the lattice space. In order to obtain the corresponding nonuniform
configuration for the staggered order parameter we consider
$\langle {\bf P}_\mu\rangle=(0,P_0,0)\delta_{\mu b}$,
and solve the equations of motion for the ${\bf n}$ fields. The
resulting local staggered magnetization reads
\beq
{\bf n}({\bf x})=(\sigma_a \cos{({\bf Q}\cdot{\bf x})},\;\sigma_b,\;
\sigma_c \sin{({\bf Q}\cdot{\bf x}})),
\label{helix}
\eeq
with $\sigma_c=-\sigma_a(Q^2+m_a^2)/{\tilde P}_0Q$ and
\beq
2Q^2={\tilde P}_0^2-m_a^2-m_c^2+\sqrt{({\tilde
P}_0^2-m_a^2-m_c^2)^2-4m_a^2m_c^2},
\label{Q}
\eeq
for ${\tilde P}_0^2>m_a^2+m_c^2$, where ${\tilde P}_0\equiv-
\rho_s{\tilde\lambda}P_0$. The equation of motion for the
longitudinal component yields $n_b^2=1-n_a^2-n_c^2$, and that
determines $\sigma_b$. The static configuration for the
staggered order parameter is nonuniform and its transverse
components, $n_a$ and $n_c$, rotate in a plane perpendicular to
the $b$ direction. Therefore the staggered order parameter
becomes {\it helicoidal}, as shown in Fig.\ \ref{Fig-helicoid}.
Such a magnetic texture arises as a result of the competition
between the anisotropies, $m_a$ and $m_c$, which favor the
staggered magnetization along the $b$-axis, and $\langle
P_\mu^\alpha\rangle$, which supports a rotation of the {\bf n}
field in the $ac$ plane. Since $\langle P_\mu^\alpha\rangle\sim x$ is
small, the resulting spin configuration has small components in
the $ac$ plane, as shown in Fig.\ \ref{Fig-helicoid}.

The local magnetization will precess around the $c$-axis, see
Fig.\ \ref{Fig-helicoid}, because $
\langle {\bf L}({\bf x})\rangle=(1/2)\left(\langle{\bf
n}({\bf x})\rangle\times{\bf D}_+\right)$,
but the {\it total space integrated} magnetization, which is measured
by magnetic susceptibility measurements, is still along the $c$-axis
$$
{\bf M}=\frac{1}{V}\int\rmd^2{\bf x}\langle {\bf L}({\bf x})\rangle=
(0,0,\frac{\sigma_b D_+}{2}),
$$
with $\sigma_b^2=1-\sigma_a^2/2-\sigma_c^2/2$. Since the magnetic
susceptibility is obtained from the {\it total} magnetization,
${\bf M}$, we conclude that the nonuniform configuration in Eq.
(\ref{helix}) is consistent with the experiments of Lavrov {\it et
al.} \cite{Lavrov}, which indicate that the Cu$^{++}$ spins remain
confined to the orthorhombic $bc$ plane throughout the SG phase,
thus yielding a featureless $\chi_a$ \cite{Lavrov}.

The new helicoidal magnetic structure gives rise to IC peaks in
neutron scattering at wave vector ${\bf Q}=(0,Q,0)$, corresponding
to the inverse helix pitch. Assuming that in the SG phase Eq.
(\ref{Q}) holds with renormalized magnon gaps, $M_a$ and $M_c$, we
obtain $Q^2=\tilde{P}_0^2-M_c^2\sim x^2-M_c^2$, because
$M_a(x>x_{AF})=0$. This expression allows us to understand both
the linear dependence of the incommensurability at higher doping
$x$, when $M_c$ is negligible, as well as a possible deviation
from linearity in a region of sizable $M_c$ ($x\rightarrow
x_{AF}$). In fact, neutron scattering experiments do exhibit such
deviation from linearity at $x=0.024$ \cite{matsuda}.

We can now make an important new theoretical prediction about the
behavior of the IC peaks in the presence of an applied magnetic field
${\bf B}$ $\perp$ to the {\cuoo} planes. Since in this case the XY gap
softens \cite{Papanicolaou}, the incommensurability is expected to reapproach
the linear behavior, $Q\sim x$, and is thus larger than for the zero
field case. In the stripe picture, on the other hand, the peak
positions remain unchanged \cite{bella-lake} and thus a perpendicular
magnetic field can be used in order to decide between these two
scenarios.

Finally, we provide an independent estimate for the critical doping
concentration $x_{AF}^{th}$ at which such helicoidal instability takes
place, from the behavior of the dipolar mass ${\cal M}_{bb}$ at $T=0$.
Assuming that ${\cal M}_{bb}$ vanishes at $x_{AF}^{th}$ we find
(recall that $k_F=k_F(x)$)
\begin{eqnarray}\label{xc1}
\mu_{bb}^2&=&{\tilde\kappa}\left(\frac{{\tilde\lambda}}{\rho_s}
\right)^2\frac{M_c}{3\pi}\left(\frac{ck_F}{M_c}\right)^3
\left\{1-\left[1-2\left(\frac{M_c}{ck_F}\right)^2
\right]\right.\nonumber\\
&\times&\left.\sqrt{1+\left(\frac{M_c}{ck_F}\right)^2}-2
\left(\frac{M_c}{ck_F}\right)^3\right\}.
\end{eqnarray}
In order to estimate $x_{AF}^{th}$ we use the bare value for the
mass of the XY mode at the critical point,
$m_c\sim
5\cdot10^{-2}$, the spin-wave velocity $c=2S\sqrt{2}$, and for
lightly doped LSCO ($x\sim 0.01$), we have $ck_F/M_c\sim 5$, with
$k_F=\sqrt{\pi x}$. Thus, in Eq. (\ref{xc1}) we can consider
$M_c/ck_F$ as a small parameter to obtain
$x_{AF}^{th}=\pi(2\mu_{bb}^2\rho_s^2/c{\tilde \kappa}{\tilde
\lambda}^2)^2=0.0206 \simeq 2.1\%$, since ${\tilde\kappa}\sim1.1$,
${\tilde\lambda}\sim 1$, $\rho_s\sim S^2$, and $\mu_{bb}\sim 1$
\cite{SS}, consistent with the vanishing of the DM gap, and in
agreement with experiments.

{\it Conclusions}.$-$
We propose a description of lightly doped LSCO in terms of a
dipolar frustration model of a canted antiferromagnet. For
$x<x_{AF}$ the dipole-magnon interaction leads to a reduction
of the DM and XY gaps, as well as of the spin stiffness, and the
qualitative features of the magnetic susceptibility remain the same as
in the undoped compound \cite{marcello}, in agreement with experiments
\cite{Lavrov}. The robustness of the AF state stems from the DM and XY
anisotropies, and for $x_{AF}<x<x_{SC}$ the ground state is
unstable towards a helicoidal phase where the local WF moment
precesses around the $c$-axis (see Fig.\ \ref{Fig-helicoid}). Such
helicoidal magnetic structure gives rise to two IC peaks along the $b$
direction, as observed in neutron scattering \cite{matsuda}, while
yielding a featureless $\chi_a$ \cite{Lavrov}. The incommensurability is
expected to scale linearly with doping, but to deviate from linearity
as $x\to x_{AF}^+$. We also predict that such linearity is recovered
once a perpendicular magnetic field is applied. Our theory provides,
for the first time, a consistent description which reconciles neutron
scattering, Raman, and magnetic susceptibility measurements in
lightly doped LSCO, and inserts the latter into a wider class of
Lifshitz helimagnets such as Ba$_2$CuGe$_2$O$_7$ and
K$_2$V$_3$O$_8$ \cite{bogdanov}.

The authors acknowledge invaluable discussions with Y. Ando, L.
Benfatto, A. Lavrov, and O. Sushkov.


\end{document}